# An approach for improving the concept of Cyclomatic Complexity for Object-Oriented Programming

Ankita
Thapar University
Patiala, India
ankitagarg60@gmail.com

*Abstract*— Measuring software complexity plays an important role to meet the demands of complex software. The cyclomatic complexity is one of most used and renowned metric among the other three proposed and researched metrics that are namely: Line of code, Halstead's measure and cyclomatic complexity. Although cyclomatic complexity is very popular but also serves some of the problems which has been identified and showed in a tabular form in this research work. It lacks the calculation of coupling between the object classes for object oriented programming and only calculates the complexity on the basis of the conditional statements. Thus there is requirement to improve the concept of cyclomatic complexity based on coupling. Paper includes the proposed algorithm to handle the above stated problem.

## I. Introduction

The primary focus of researchers, technicians and software engineers is to get quality software and for that reason it is very necessary to measure the software complexity. Mainly three software metrics are used to measure software complexity LOC, Halstead's measure of complexity and McCabe's cyclomatic complexity. The main motive of this paper is to measure the exact complexity of software. Complexity of software affects various factors like maintenance cost, testing effort, quality, man-hours etc. McCabe cyclomatic complexity measure does not calculate the exact complexity of software. It only calculates the software complexity via counting the number of decisional statements from the source code or via number of independent paths through the control flow graph and that calculated number should be in the range of 1-10 and range specifies that software is error free. So, from that cyclomatic complexity number some prediction can be made in advance about the quality of software. Cyclomatic complexity does not calculate the complexity of software if there is interaction between object classes for object oriented programming. That coupling complexity also needed to be in consideration because a change in one affects other. So, in this paper an algorithm is purposed to measure the interaction between object classes based on all the unique external class references.

## II. Related Work

Loc metric considered as a traditional metric which counts the number of lines form the source code and it was not considered as an adequate metric because if a source code contains 500 lines consisting of 100 decision statements and which may contain million of paths then with Loc metric it is possible to measure and test only small proportion [1]. After that in 1977 Halstead's measure of complexity was purposed which was used to calculate the software complexity by counting the number of unique operators and operands [2] but it ignores the complexity from the control flow graph and put the same emphasis if source code contain operands and operators with branches or not [3]. So, for that McCabe's cyclomatic complexity metric was purposed which includes the complexity from the control graph. This software complexity metric was used to measure the software complexity by three methods. In first method: generate the control flow graph from the source code and then count the number of nodes and edges from that source code. Second method is considered as a simplest method and this method simply counts the number of decision or conditional statements directly from the source code's and that's why also called conditional complexity [4] and last method counts the number of regions form the control flow graph. It is considered that more complex software has more number of errors and after that more effort will be required to correct these errors and then it becomes difficult to change the software in the future. McCabe's cyclomatic complexity was considered as a strongest metric among three. Our main purpose is to measure the appropriate software complexity. But McCabe's cyclomatic complexity metric failed to measure the exact software complexity if the interaction between different classes is higher. So, the concept of object coupling needs to be included in the concept of cyclomatic complexity. Coupling factor also needs to be introduced in McCabe cyclomatic complexity which lags behind as it does not consider basic elements like class, polymorphism, encapsulation etc. [5] Coupling should be taken care of because it measures the degree of interdependence between two modules or object classes means change in one object class affects another object classes [6] . Two or more classes can be either loosely coupled or highly coupled based on the number of ways they are coupled. If two classes are coupled in more than one way then they are highly coupled classes else loosely coupled.
If coupling is not considered while calculating the complexity it will lead to some pitfalls in the quality factors of the program. The factors that are affected by this includes reliability [7], reusability, understand ability, cost, testing effort, fault prediction, quality, modifiability, maintainability.
There are two types of interactions: one is between the object classes and another one is between the modules with in the class i.e. intra class interaction. But in this paper the focus is on inter-class coupling.

### III. PROBLEM STATEMENT AND RELATED SOLUTION

McCabe cyclomatic complexity is not considering or measuring the exact software complexity means if there is interaction between two or three object classes in software then it does not calculates that complexity. There is a need to consider that object coupling complexity in McCabe cyclomatic complexity.
Coupling can occur among object classes through different methods like:
Field accesses, through methods calls, Inheritance, Arguments, Return types, Exception, instruction type.

There are some basic rules that can be used to measure McCabe's cyclomatic complexity number.

1. Calculate the number of if/then, else if but do not count the else statements in the program.
2. Find the switch statement and count the total of the cases in the program but do not count the default in the program.
3. Calculate all the loops like for, while and do-while statements and also all the try/catch statements in the program.
4. Count conditional operator && and || operator and also ternary operators like ?: from the expression [8].

Now, add one to the numbers from the previous step numbers. And now the algorithm for McCabe cyclomatic complexity

#### A. Algorithm 1
Find McCabe's cyclomatic complexity

Input: Java File
Output: Cyclomatic complexity
/*keyword is an array of all the decision making keywords which affect cyclomatic complexity, cyclomatic is the variable that holds the cyclomatic complexity
Initialize it with 0*/

1. Read the input file
2. foreach java file Read it with bufferedreader line by line
       while (buffer! = NULL)
           for each line divide it into tokens with stringTokenizer

           if (token equals keyword[i])
               cyclomatic ++;
                   break;
           endif
       endfor
   endwhile
endfor
3. Final cyclomatic complexity will be cyclomatic +1.

There are three methods used to measure the cyclomatic complexity of software. In First method, firstly draw the control flow graph from the source code's and then counts the number of edges and nodes from control flow graph. So, number of independent paths generated from graph is equal to cyclomatic complexity number. Second method is one of simplest method. This method simply counts the number of conditional statements from the source code's directly based on above four rules and last method counts the number of regions from the control graph. From all three methods we get the same Cyclomatic complexity number. This calculated cyclomatic complexity number should be in range of 1 to 10 and only then software is considered as risk free software. If the same lies in the range of 10-20 then it is considered as a target of moderate risk. 30-40 range of cyclomatic number makes module highly risky and the range exceeding 40 exempt it from the candidate of testing [9]. Based upon some data it has been proven that higher will be the cyclomatic complexity number lower will be the quality of software.

So, from above all discussion, McCabe's cyclomatic complexity metric only considers complexity from control graph or conditional statements and not considers complexity is there interaction or coupling between two object classes.
Now, algorithm is purposed based on Field accesses through methods calls, Inheritance, Arguments, Return types, Exception, instruction type.

#### B. Algorithm 2
Purposed Algorithm for Coupling between object classes

Input: Any class file
Output: Total coupling in the class file
/*Eff is set for efferent coupling(set is java collection which holds only unique values), i is the index used, Interface contains the number of interfaces, Field contains the number of fields, Method contains the number of methods, filename is the name of input file, argumentstype is the type of argument, exceptiontype is the type of exception. Here class is being iterated using an API bcel.jar*/

1. Select a java class file and parse it using bcel API. //bcel parses it and creates objects of class
2. Set: filename = inputfilename;
3. Traverse the class.
4. foreach Interface ϵ class do
               registerCoupling (Interface);
       endfor

5. foreach Field ∈ class do
                    registerCoupling (Field.returntype);
    endfor
6. foreach Method ∈ class *do   //Identify different classes based on arguments, returntypes*
             registerCoupling (Method.returntype);
             registerCoupling (Method.argumentstype);
             registerCoupling (Method.exceptiontype);
             registerCoupling (Method.localargumentstype);
         foreach MethodInstruction
    */* if conditions can be determined by using api bcel.jar and calling its api*/*
           if (instruction instanceof  LocalVariableInstruction)
              registerCoupling (LocalVariableInstruction type);
           endif
           if (instruction instanceof  ArrayInstruction)
              registerCoupling (ArrayInstruction type);
           endif
           if (instruction instanceof  FieldInstruction)
              registerCoupling (FieldInstruction type);
           endif
           if (instruction instanceof  InvokeInstruction)
              registerCoupling (InvokeInstruction type);
           endif
           if (instruction instanceof  INSTANCEOF)
              registerCoupling (INSTANCEOF type);
           endif
           if (instruction instanceof  CHECKCAST)
              registerCoupling (CHECKCASTtype);
           endif
       endfor
    endfor
endfor
7. Return Eff.size();

Methods
registerCoupling (Type t){
     registerCoupling (className(t))
}
className(Type t){
        if (t is premetive)
               return;
        endif
        if (t is array type)
               return;
        endif
               return name of type t ;
/**in java it can be t.tostring()**/
               }
registerCoupling (String classname){
         if (classname is javaclass or classname is filename)
               return;
         endif
         /**Add to set**/
         Eff.add(classname)
}

The algorithm here uses a library BCEL (byte code engineering library) which helps to understand and read the byte code of java file. It has API's which can read the class file line by line and gives the return type, no of interfaces etc. Here a class HashSet which will save all the unique external references to other classes is used. Since in java Hashset which is implementation of Set store only unique values. This data structure proved very helpful the intentions were to get unique references.

Also there is a method which has been frequently called throughout the algorithm is registerCoupling (). This method takes the class type or class name as input parameter and adds it to the set. This method has been overloaded, one taking a String parameter and other taking Type (BCEL API's class) which calls other method className(Type t) (reference: See methods section at the end of algorithm) which returns the name of the class and make sure that this is not a primitives type. Only non primitive's types are considered and in registerCoupling (String name) method, check whether it is a java api class or not if not then add it to the set.

So, the algorithm starts with a java .class file (which is basically byte code) being passed as input and its name is stored in a variable. The purpose of this variable is, as CBO is number of unique external class references. So, this will not be taken in to consideration as a class name.

Since, CBO is total number of unique references, whether it is inheriting some other class or some local variable of other class type or exceptions or return type of a method or instance of other class algorithm tries to cover all the permutations.

In step 4 and 5 all the interfaces and all the class level fields respectively are considered and added their class to the set by calling method registerCoupling.

In step 6 it has started reading all the methods of the class. Firstly return type, exception type and argument types of this method are determined and these types are added to set again by calling registerCoupling method (). Now, method is traversed line by line and each line has various type of instructions. These can be anything like a local variable is declared etc. For these instructions BCEL has provided various classes such as LocalVariableInstruction (i.e. a local variable has been created) INSTANCEOF (i.e. whether a field or object is instance of some other class) and these classes are again added to set by calling registerCoupling method.

Finally, the length of this set is the final CBO of the input class.

Now, Final McCabe cyclomatic complexity number should be NewCyclomatic complexity = Cyclomatic Complexity number + Coupling between object classes.

IV. RESULTS

The table I shown below consists of various classes which have been taken from BCEL (Byte code engineering library) API. This API has been used in proposed algorithm. Firstly cyclomatic complexity number is calculated according to cyclomatic complexity concept (i.e. conditional statements) in first column. Along with a prediction can be made from this number about the quality of software, about maintenance cost, reliability, understanability etc. But it does not calculate the exact software complexity. There is a need to consider other factors to calculate the exact complexity of software so that an accurate prediction and assumption can be made from the exact complexity number. Cyclomatic number does not consider any complexity when a code interacts with other code e.g. Class A is calling some function of class B. For these interactions then for that there is a another factor i.e. CBO (complexity between objects). CBO is nothing but just the number of unique references a class has to other external classes by any means say it can inherit other class or some of its method has a return type of some external class etc.

Other possibilities will be discussed in the algorithm for calculating CBO e.g. Class A inherits Class B and it has some function whose return type is Class C. In that case CBO for Class A will be 2.

So, in second column coupling between object classes is calculated based on all the unique external class references because cyclomatic complexity metric does not include the complexity if there is an interaction between object classes. Now, table includes various classes in which proposed algorithm has been applied and a new approach cyclomatic number plus coupling between classes has been applied for improving the concept of cyclomatic complexity.

Table I: New cyclomatic complexity of different API classes

| S.No | Classes | Cyclomatic complexity number | Coupling between object classes | New Cyclomatic complexity number |
|---|---|---|---|---|
| 1 | org.apache.bcel.classfile.AccessFlags | 4 | 0 | 4 |
| 2 | org.apache.bcel.classfile.Attribute | 18 | 21 | 39 |
| 3 | org.apache.bcel.classfile.AttributeReader | 1 | 2 | 3 |
| 4 | org.apache.bcel.classfile.ClassFormatException | 1 | 0 | 1 |
| 5 | org.apache.bcel.classfile.ClassParser | 17 | 6 | 23 |
| 6 | org.apache.bcel.classfile.Code | 20 | 7 | 27 |
| 7 | org.apache.bcel.classfile.CodeException | 4 | 5 | 9 |
| 8 | org.apache.bcel.classfile.Constant | 13 | 17 | 30 |
| 9 | org.apache.bcel.classfile.ConstantClass | 1 | 5 | 6 |
| 10 | org.apache.bcel.classfile.ConstantCP | 1 | 2 | 3 |
| 11 | org.apache.bcel.classfile.ConstantDouble | 1 | 4 | 5 |
| 12 | org.apache.bcel.classfile.ConstantFieldref | 1 | 2 | 3 |
| 13 | org.apache.bcel.classfile.ConstantFloat | 1 | 4 | 5 |
| 14 | org.apache.bcel.classfile.ConstantInteger | 1 | 4 | 5 |
| 15 | org.apache.bcel.classfile.ConstantInterfaceMethodref | 1 | 2 | 3 |

## V. DISCUSSION

In this section the focus is on the cyclomatic complexity. Why this is preferred over two metrics LOC and Halstead's metric, All three methods are used to measure the software complexity but only cyclomatic complexity metric is best among them because it considers the software complexity via number of decision statements or via control flow graph. Three methods are used to calculate the McCabe's cyclomatic complexity number and same result is obtained from all three methods. But, there is also a problem along with cyclomatic complexity concept. McCabe's cyclomatic complexity method is not used to calculate the software complexity if there is an interaction between two modules i.e. with-in a class or if there is an interaction between object classes means two object classes are communicating with each other and it is considered as if coupling or interaction is more then, quality of software and other attributes like reliability, modifiability, maintainability etc. get compromised. If there is some changes occur in one object class then simultaneous modification or changes occur in other class. This is a very important concept and must be in consideration. So, a solution is purposed, how can be this concept of cyclomatic complexity can be improved by calculating the coupling between object classes.

The results discussed in the table I can also be depicted through a graph shown in the figure 1 where the graph shows the actual improved cyclomatic complexity of the various BCEL API classes taken.

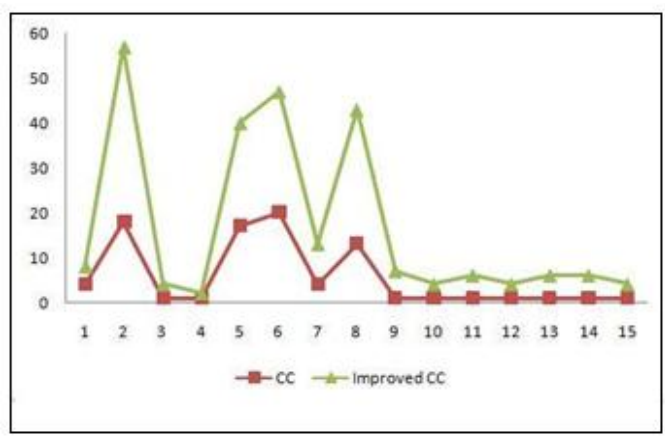

Fig.1. Improved graph for cyclomatic complexity

## VI. CONCLUSION

McCabe cyclomatic complexity measure of software complexity is one of the strongest metric among LOC, Halstead's and language independent. Till now it is used to calculate the software complexity only through control flow graph (via conditional statements) but there is a need to include the concept of coupling with cyclomatic complexity concept and for that one algorithm is purposed, how can calculate the coupling between object classes for object-oriented programming and then, used that concept in cyclomatic complexity for the improvement of cyclomatic complexity concept.